\documentclass[aps, prl, superscriptaddress, twocolumn, preprintnumbers,amsmath,amssymb,notitlepage, nobalancelastpage,10pt,longbibliography]{revtex4-2}

\usepackage{amsmath}
\usepackage{verbatim}
\usepackage{graphicx}
\usepackage{color}
\usepackage{braket}
\usepackage{yfonts}

\usepackage{dsfont}
\newcommand{\cop}{\hat{c}}
\newcommand{\cdop}{\hat{c}^\dag}
\newcommand{\Hop}{\hat{H}}

\usepackage[colorlinks=true,citecolor=blue,linkcolor=blue,urlcolor=blue]{hyperref}

\begin{document}

\title{Revealing the topological nature of the bond order wave in a strongly correlated quantum system}

\author{Sergi Juli\`a-Farr\'e}
\email{sergi.julia@icfo.eu}
\affiliation{ICFO - Institut de Ciencies Fotoniques, The Barcelona Institute of Science and Technology, Av. Carl Friedrich Gauss 3, 08860 Castelldefels (Barcelona), Spain}
\author{Daniel González-Cuadra}
\affiliation{ICFO - Institut de Ciencies Fotoniques, The Barcelona Institute of Science and Technology, Av. Carl Friedrich Gauss 3, 08860 Castelldefels (Barcelona), Spain}
\affiliation{Center for Quantum Physics, University of Innsbruck, 6020 Innsbruck, Austria}
\affiliation{Institut f\"ur Quantenoptik und Quanteninformation, \"Osterreichische Akademie der Wissenschaften, Technikerstra{\ss}e 21a, 6020 Innsbruck, Austria}
\author{Alexander\,Patscheider}
\affiliation{Institut f\"ur Experimentalphysik, Universit\"at Innsbruck, Technikerstra{\ss}e 25, 6020 Innsbruck, Austria}
\author{Manfred\,J.\,Mark}
\affiliation{Institut f\"ur Quantenoptik und Quanteninformation, \"Osterreichische Akademie der Wissenschaften, Technikerstra{\ss}e 21a, 6020 Innsbruck, Austria}	
\affiliation{Institut f\"ur Experimentalphysik, Universit\"at Innsbruck, Technikerstra{\ss}e 25, 6020 Innsbruck, Austria}
\author{Francesca\,Ferlaino}
\affiliation{Institut f\"ur Quantenoptik und Quanteninformation, \"Osterreichische Akademie der Wissenschaften, Technikerstra{\ss}e 21a, 6020 Innsbruck, Austria}	
\affiliation{Institut f\"ur Experimentalphysik, Universit\"at Innsbruck, Technikerstra{\ss}e 25, 6020 Innsbruck, Austria}
\author{Maciej Lewenstein}
\affiliation{ICFO - Institut de Ciencies Fotoniques, The Barcelona Institute of Science and Technology, Av. Carl Friedrich Gauss 3, 08860 Castelldefels (Barcelona), Spain}
\affiliation{ICREA, Pg. Llu\'is Companys 23, 08010 Barcelona, Spain}
\author{Luca Barbiero}
\affiliation{Institute for Condensed Matter Physics and Complex Systems,
DISAT, Politecnico di Torino, I-10129 Torino, Italy}
\affiliation{ICFO - Institut de Ciencies Fotoniques, The Barcelona Institute of Science and Technology, Av. Carl Friedrich Gauss 3, 08860 Castelldefels (Barcelona), Spain}
\author{Alexandre Dauphin}
\email{alexandre.dauphin@icfo.eu}
\affiliation{ICFO - Institut de Ciencies Fotoniques, The Barcelona Institute of Science and Technology, Av. Carl Friedrich Gauss 3, 08860 Castelldefels (Barcelona), Spain}

\begin{abstract}
    We investigate the topological properties of the bond order wave phase arising in the extended Fermi-Hubbard model. In particular, we uncover a topological sector, which remained elusive in previous finite-size numerical studies due to boundary effects. We first show that, for an infinite system, the bond order wave regime is characterized by two degenerate bulk states corresponding to the trivial and topological sectors. The latter turns out to be indeed characterized by an even degeneracy of the entanglement spectrum and long-range order of a string correlation function. For finite-size systems, we show that the topological sector can be stabilized by imposing a suitable border potential. This therefore provides a concrete protocol for the observation of topologically protected degenerate edge modes in finite-size systems. Furthermore, we show that the bulk of the system is characterized by exotic solitonic solutions interpolating between the trivial and topological sectors. Finally, we propose an implementation and detection scheme of this strongly correlated topological phase in a quantum simulator based on dipolar Fermi gases in optical lattices. 
\end{abstract}

\maketitle

{\textit{Introduction}}.\textemdash In the recent years, great effort has been devoted toward the study of symmetry-protected topological (SPT) phases~\cite{gu_2009, RevModPhys.82.3045, Pollman2010, xie_2012,Cooper_2019}. These exotic phases, characterized by nonlocal order parameters, escape the conventional Ginzburg-Landau theory of phases of matter~\cite{Landau1937,chiu_2016}, and their robustness with respect to local perturbations allows for applications going from metrology to quantum computation~\cite{kitaev2009,Klitzing2017}. Owing to their high level of control~\cite{lewenstein2017,Bloch_2008}, ultracold atomic systems represent an ideal platform where such intriguing states of matter can be investigated~\cite{goldman_16,Cooper_2019}. In one dimension, one of the most paradigmatic models hosting a SPT phase, the  Su-Schrieffer-Heeger (SSH) model~\cite{SSH1979}, has been realized in atomic quantum simulators~\cite{Atala2013,Meier2016} and its robustness to disorder has been probed~\cite{Meier2018}. While these experiments probed noninteracting models, the inclusion of interactions can lead to much richer phenomena~\cite{deLeseleuc2019,Sompet2021}. 

Furthermore, SPT phases can arise directly from interactions, as it is the case of the original SSH Hamiltonian in polyacetylene. There, a bond order wave (BOW) arises spontaneously from the coupling between electrons and phonons through a Peierls mechanism~\cite{peierls_96}, which can also occur in spin-boson models~\cite{gonzalez_18,gonzalez_19a,gonzalez_19b}. 

Interestingly, similar BOW phases also appear in interacting single-species systems, induced by frustration between competing orders~\cite{Aligia2007,Otsuka2005,Tincani2009,Hallberg1990,Schmitteckert2004,Mishra2011,mondal2022realizing,Nakamura99, Nakamura2000,Jeckelmann2002,Sengupta2002,Sandvik2004,Zhang2004,Tam2006,Glocke2007,Ejima2007,Barbiero2014,Hallberg1990,Somma2001,Sato2011,Kumar2010,Mishra2013}, including  strongly-correlated electrons, quantum magnets, or ultracold atomic systems. Although the insulating nature and effective dimerization of these systems have been very carefully characterized, their topological nature still need to be unveiled. The latter requires an accurate analysis of such many-body interacting systems, as the sole presence of a spontaneous dimerization, i.e., a local order parameter, does not directly translate into a nontrivial topology.

In this Letter, we reveal and characterize the SPT nature of such BOW phases arising from frustration in the presence of chiral symmetry~\cite{Hallberg1990,Schmitteckert2004,Mishra2011,mondal2022realizing,Nakamura99, Nakamura2000,Jeckelmann2002,Sengupta2002,Sandvik2004,Zhang2004,Tam2006,Glocke2007,Ejima2007,Barbiero2014,Hallberg1990,Somma2001,Sato2011,Kumar2010,Mishra2013}. Moreover, we propose a realistic implementation and detection scheme for the realization of the frustration-induced topological BOW phase with dipolar gases in optical lattices. Our proposal allows us to go beyond the experimental simulation of noninteracting SPT phases, promising to access both bulk and edge physics of a strongly correlated topological phase with richer phenomenology.

More specifically, we focus our analysis on the one-dimensional (1D) extended Fermi-Hubbard (EFH) model. We demonstrate that, in the case of an infinite chain, two exactly degenerate BOW ground states occur. These states are invariant under chiral and bond-inversion symmetries, and correspond to the topological and the trivial ground states of the interacting SSH model~\cite{Gurarie2011,Manmana2012,Yoshida2014,Wang2015,Ye2016,Sbierski2018,Barbiero2018,Le2020}. As indeed required by SPT phases, we find that the topological state is characterized by the long-range order of a specific nonlocal string correlator ~\cite{Nijs1989,Barbiero2013,Montorsi2017} and by an even degeneracy of the entanglement spectrum (ES) \cite{Zaletel2014,Pollman2010}. Furthermore, we show that, in a finite size system, the topological sector of the BOW can be stabilized by means of a suitable local pinning. In this case, the \emph{bulk-edge} correspondence of SPT phases translates into the presence of gapless spin edge modes that were not observed in previous finite-size studies. Exemplary to the rich phenomenology of the system, we find that further spin bulk excitations create solitonic structures interpolating between the topological and trivial sectors of the BOW. 

Finally, we propose an experimental setup based on erbium magnetic atoms trapped in a 1D optical lattice. Impressive steps forward have been achieved in ultracold systems made of magnetic atoms with large dipolar momenta to simulate large and nonlocal interactions~\cite{Norcia2021nof}. The peculiar nonlocal dipolar repulsion has made possible the experimental study of exotic states of matter, such as droplet liquids \cite{Chomaz2016,Schmitt2016} and supersolids \cite{Boettcher2019,Tanzi2019a,Chomaz2019, Norcia2021}, as well as the investigation of ergodic behaviors \cite{Tang2018}. Furthermore, it has been also experimentally demonstrated that, when trapped in a lattice, magnetic atoms mimic the physics of extended Hubbard Hamiltonians \cite{Trefzger2011,dePaz2013,Dutta2015,Baier2016,Lepoutre2019,Patscheider2020}. As we reveal, such a setup allows, on one side, to have sizable nonlocal repulsion required to achieve the BOW regime and, on the other, to perform very accurate measurements of density distribution and string correlators through a quantum gas microscope~\cite{Endres200,Hilker484,Haller2015,Mazurenko2017}.

{\textit{Extended Fermi-Hubbard model}}.\textemdash The EFH model describes a chain of length $L$ where $N$ spinful fermions, labeled by $\sigma=\uparrow,\downarrow$, interact through contact and nearest-neighbor (NN) repulsion. The Hamiltonian modeling such a system reads as
\begin{equation}
 \label{eq:hamiltonianfree}
\begin{split}
\Hop=&-t\sum_{\langle ij\rangle, \sigma}(\cdop_{i,\sigma}\cop_{j,\sigma}+\textrm{H.c.}) +U\sum_{i=0}^{L-1}\hat{n}_{i,\uparrow}\hat{n}_{i,\downarrow}+\\
&+V\sum_{\langle ij\rangle} \hat{n}_{i}\hat{n}_{j},
\end{split}
\end{equation}
where $t$ parametrizes the NN hopping, $U$ accounts for the on-site Hubbard interaction, and $V$ describes the repulsion between fermions in NN sites. Here, we restrict our investigation to the case where both $N$ and the total magnetization $\hat{S}_z\equiv\sum_{i}(\hat{n}_{i,\uparrow}-\hat{n}_{i,\downarrow})/2$ are conserved and, unless specified, we consider the half-filled case with $N=L$ and $\hat{S}_z=0$. We emphasize that, for spinful fermionic systems, both charge and spin degrees of freedom have to be considered. More precisely, we refer to a gapped charge or spin sector when the system has to pay a finite energy for adding/removing an up-down pair, or flipping a single fermion, respectively.
\begin{figure}[t]
\includegraphics[width=
\columnwidth]{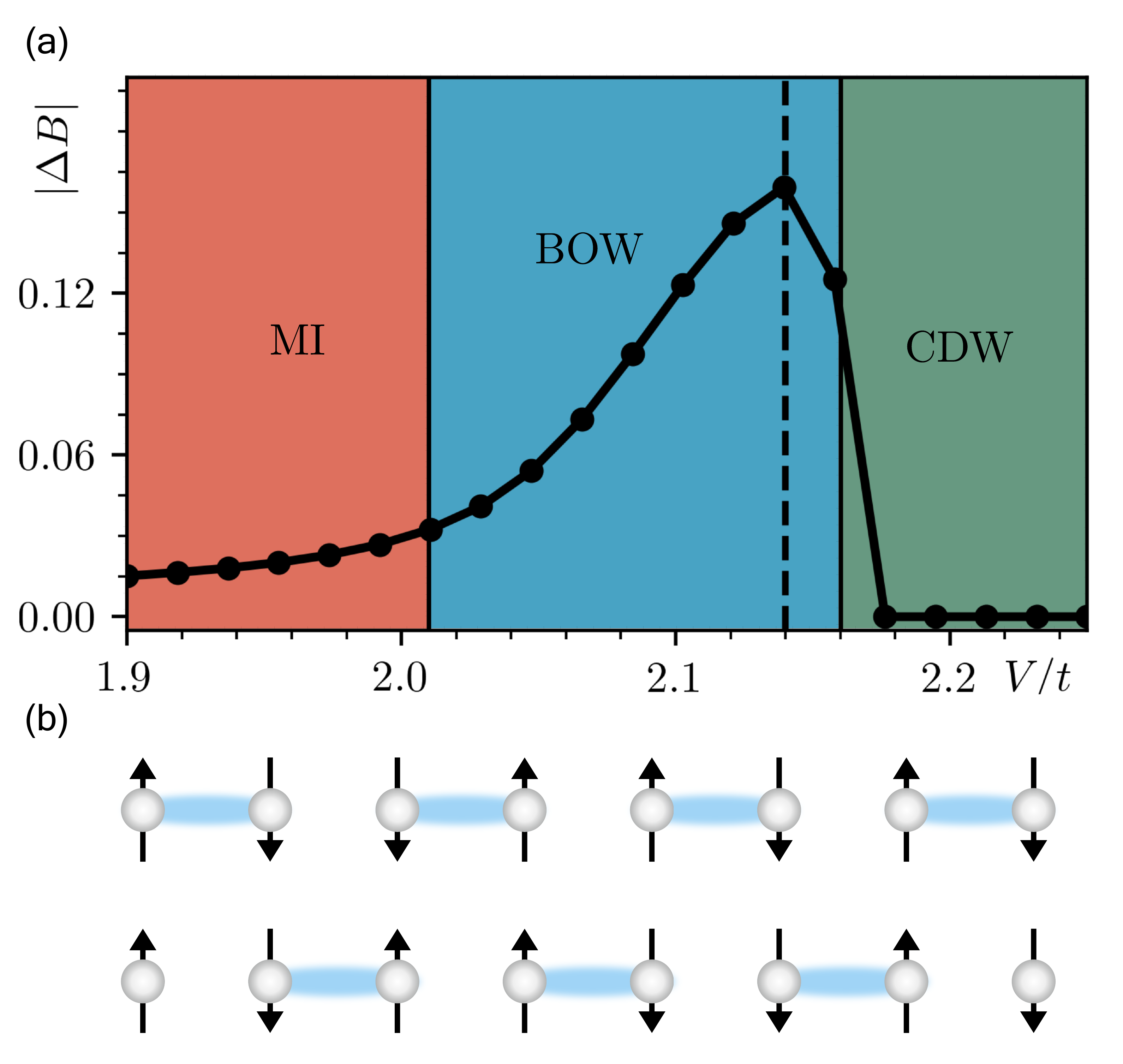}
\caption{(a) $|\Delta B|$ at fixed $U=4t$ and different values of $V/t$. We use the iDMRG algorithm with a two-site unit cell and a large bond dimension $\chi= 3000$. The dashed line $V/t\simeq 2.14$ corresponds to the maximum value of $\Delta B$. (b) Sketch of the spontaneous dimerization in the BOW phase, with bonds corresponding to a large $|\hat{B}_i|$ depicted in blue. The upper chain corresponds to the trivial case, while the lower chain is topologically nontrivial and exhibits edge states under open boundary conditions.}
\label{fig:pd}
\end{figure}

{\textit{Phase diagram}}.\textemdash When $U$ dominates the system is a Mott insulator (MI) with a finite charge gap and short-range antiferromagnetic order. Alternatively, for large $V$ the system has a fully gapped charge density wave (CDW), characterized by an effective antiferromagnetic order, with alternating empty and doubly occupied sites. In the strongly interacting limit $U,V\gg t$ there is a direct transition between these two phases at $U=2V$. However, when $V$ and $U$ compete and are comparable to the hopping amplitude $t$, frustration effects induce a third phase, the fully gapped BOW~\cite{Nakamura99, Nakamura2000,Jeckelmann2002,Sengupta2002,Sandvik2004,Zhang2004,Tam2006,Glocke2007,Ejima2007,Mishra2011,Barbiero2014}. Such a phase is characterized by a uniform distribution of particles, as in MIs, accompanied by a spontaneous dimerization that leads to a staggered expectation value of the bond operator $\hat{B}_{i}=\frac{1}{2}\sum_{\sigma}(\cdop_{i,\sigma}\cop_{i+1,\sigma}+\textrm{H.c.})$ captured by $\Delta B_{i}\equiv \langle \hat{B}_{i}-\hat{B}_{i+1}\rangle$. Interestingly, including dipolar interactions beyond the NN term sensibly enlarges the range of parameters where the BOW can be found \cite{Barbiero2014}.

Here, we complement previous analyses in finite systems by performing infinite density matrix renormalization group (iDMRG) calculations~\cite{tenpy} where boundaries do not play any role, thus allowing one to study only the properties of the bulk of the system.
Figure~\ref{fig:pd}(a) shows $|\Delta B|\equiv |\Delta B_{L/2}|$ as a function of $V$ for ${U=4t}$: While the BOW-CDW transition can be accurately determined at $V_{\text{BOW-CDW}}\simeq 2.16t$, the Berezinskii-Kosterlitz-Thouless nature of the MI-BOW transition makes it challenging to derive the transition point accurately (see Supplemental Material \cite{SM}). Previous finite-size extrapolations of the thermodynamic limit yield $V_\textrm{MI-BOW}\simeq 1.88t$~\cite{Ejima2007,Sengupta2002}. However, the more recent study of Ref.~\cite{Dalmonte2015}, based on a gap-scaling analysis in finite systems, results in a considerably larger value $V_\textrm{MI-BOW}\simeq 2.08t$. Here, by working directly in the thermodynamic limit, we refine such predictions, and obtain $V_\textrm{MI-BOW}\gtrsim 2.01t$.

 \emph{Degeneracy of the BOW phase}.\textemdash Our iDMRG calculations allow one to identify an exact bulk degeneracy between the two ground states. Such equivalent bulk solutions correspond to the two bulk sectors of the spontaneously symmetry-broken BOW with $\pm |\Delta B|$ [Fig.~\ref{fig:pd}(b)]. Notice that these two degenerate ground states correspond to two effective lattice dimerizations, which are reminiscent of the two possible static dimerizations in the SSH model. Indeed, the behavior of the parity operators in the BOW phase~\cite{Barbiero2013} is the same as in the SSH model with on-site repulsion~\cite{Barbiero2018}. In what follows, by characterizing the topology of the BOW in the bulk, we confirm that the system realizes a SPT phase protected by chiral and inversion symmetry. 
 
\begin{figure}[t]
\includegraphics{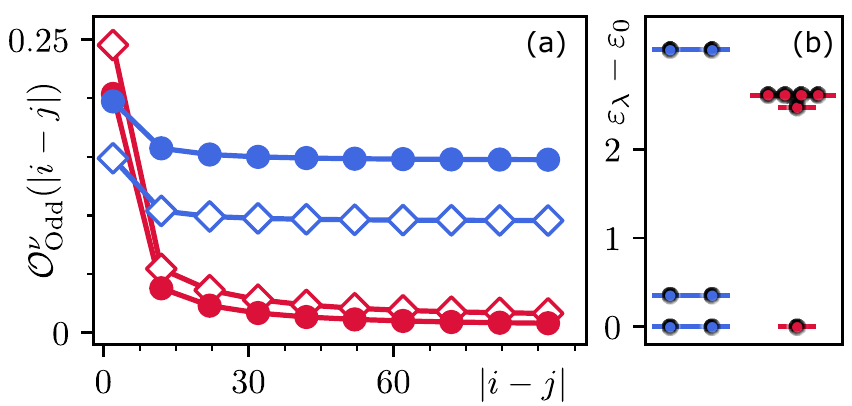}
\caption{Topological properties of the BOW at half filling for $U=4t$ and $V=2.14t$ obtained with iDMRG.  Red (blue) colors are used for the trivial (topological) sectors. (a) Long-range behavior of the odd string order parameters. Solid circles (open  squares) are used for the spin (charge) strings. (b) ES energies $\varepsilon_\lambda \equiv -\log S_\lambda^2$. }
\label{fig:ent_so}
\end{figure}

 {\textit{Characterization of the topological bulk sector}}.\textemdash For one-dimensional interacting systems, the presence of a SPT phase is signaled by a nonvanishing value of a nonlocal string order parameter~\cite{Nijs1989,Barbiero2013,Montorsi2017} that can be measured in ultracold atomic systems with a quantum gas microscope~\cite{Endres200,Hilker484,Haller2015,Mazurenko2017}. In particular, the topological nature of SSH-like chains is captured by the long-range order of the following string correlator~\cite{Anfuso2007,Wang2013,Fraxanet2021}, 
\begin{equation}
\begin{split}
& {O}^\nu_{\text{odd}}(|i-j|) =\left|4 \langle \hat{S}^\nu_{2i+1}\exp\left[i\pi\sum_{k=2i+2}^{2j-1} \hat{S}^\nu_k\right]\hat{S}^\nu_{2j}\rangle   \right|,
\end{split}
\label{eq:string}
\end{equation}
where $\nu=s,c$ denotes the spin and charge sectors. Due to the fully gapped nature of the BOW phase, we calculate the string correlator~\eqref{eq:string} both in the spin sector, where $\hat{S}^s_i=\frac{1}{2}(\hat{n}_{i,\uparrow}-\hat{n}_{i,\downarrow)}$ and in the charge sector, where $\hat{S}^c_i=\frac{1}{2}(\hat{n}_i-1)$. Figure~\ref{fig:ent_so}(a) shows the spin and charge strings of the degenerate ground states. A proper scaling of these quantities allows one to infer the limit of ${O}^\nu_{\text{odd}}\equiv\lim _{|i-j|\to \infty}{O}^\nu_{\text{odd}}(|i-j|)$: ${O}^s_{\text{odd}}$ and ${O}^c_{\text{odd}}$ are finite for the gapped topological spin and charge sectors and vanish for the topologically trivial phase. Since the two bulk ground states are identical up to a translation of one site, the even string order ${O}^\nu_{\text{even}}(|i-j|) =\left|4 \langle \hat{S}^\nu_{2i}\exp\left[i\pi\sum_{k=2i+1}^{2j} \hat{S}^\nu_k\right]\hat{S}^\nu_{2j+1}\rangle   \right|$ has the opposite property: It is finite in the trivial sector and vanishing in the topological one. Nevertheless, as our goal is to characterize both bulk and edge topological properties, ${O}^\nu_{\text{odd}}$ is the proper observable to predict the appearance of edge states in finite-size systems.

Moreover, the degeneracy of the ES also allows one to characterize the topology of the degenerate ground states~\cite{Pollman2010,Zaletel2014}: In 1D, it has been shown that, under the preservation of their protecting symmetries, SPT phases exhibit an even degeneracy of the ES, and therefore phases with the same ES degeneracy can be connected adiabatically and are thus topologically equivalent. We therefore compute the ES $S_\lambda$, given by the eigenvalues of the reduced density matrices for a bipartite cut of the infinite chain. Figure~\ref{fig:ent_so}(b) shows the ES for the two degenerate iDMRG ground states of the BOW: The BOW is either a trivial phase with a lack of even degeneracy of the ES or a topological phase with an even degenerate spectrum, as in the dimerized SSH-Hubbard model~\cite{Yoshida2014,Wang2015,Ye2016,Barbiero2018}. The latter is consistent with the previous string order analysis.

\begin{figure}[t]
\includegraphics[]{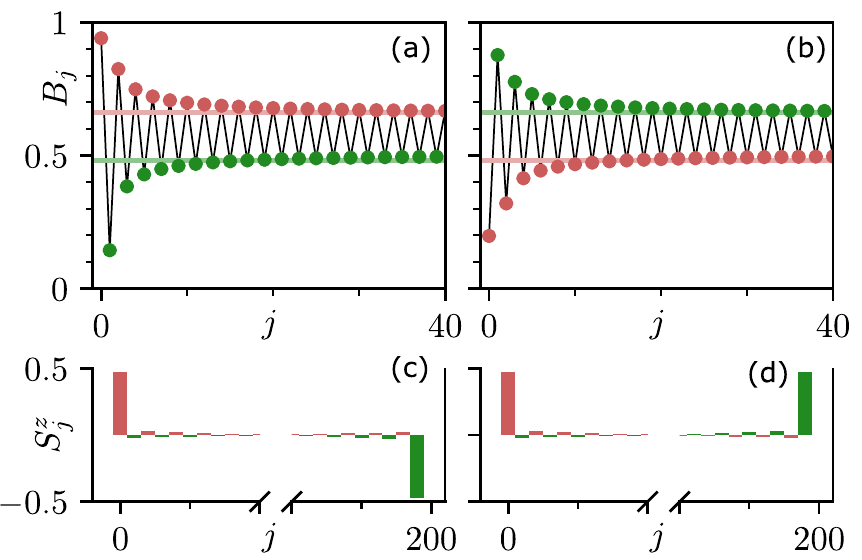} \caption{Finite DMRG results at half filling for $U=4t$, $V=2.14t$, $L=200$, and $\chi_\text{max}=1200$. Red (green) points are used for the even (odd) bonds or sites. (a),(b) Expectation value of the bond operator in the BOW phase exhibiting the trivial (a) and topological (b) staggered patterns. Only the first sites on the left part of the chain are shown, as the bond profile is symmetric with respect to its center. Solid lines represent the iDMRG value with $\chi_\textrm{max}=3000$. (c),(d) Local polarization of two degenerate topological solutions corresponding to the bond staggerization of (b) for $S_z=0$ (c) and $S_z=+1$ (d).}
\label{fig:real_space}
\end{figure}

{\textit{Topological edges in finite-size systems}}.\textemdash For a finite-size system, border effects break the degeneracy of the two ground states and the topological dimerized pattern turns out to be an excited state for open boundary conditions: In the bulk, the fermions always tunnel to the left/right site with an effective hopping strength $t(1\pm |\Delta B|)$ but, at the edge of the chain, the system is forced to select the most favorable hopping configuration, namely the one given by $t(1+|\Delta B_0|)$. Therefore, previous finite DMRG studies of Eq.~\eqref{eq:hamiltonianfree} only focused on the state related to the trivial topology. 
We now show how such a BOW phase can be stabilized in the presence of edges with finite DMRG. In order to select a given dimerization, we use a local pinning field that fixes the bond pattern at the borders of the chain~\cite{SM}. Figures~\ref{fig:real_space}(a) and \ref{fig:real_space}(b) show the two staggered bond patterns obtained by varying the sign of the pinning field, that correspond to the trivial and topological BOW phase, respectively. Figure~\ref{fig:real_space}(c) shows the spin-polarized edge states only appearing in the topological sector. As these edge states couple weakly with the bulk (see Supplemental Material~\cite{SM}), we can approximate the reduced density matrix of the edges by the product state wave function $\ket{\Psi}_{\rm{edges}} =\ket{\cdot}_L\ket{\cdot}_R$, where $\ket{\cdot}_{L(R)}$ represents the quantum state of the first (last) site of the chain. 
Let us now discuss the degeneracy of such an edge state manifold. In the $\hat{S}_z=0$ sector, the system has two degenerate topological ground states corresponding to $\ket{\downarrow}_L\ket{\uparrow}_R$ and $\ket{\uparrow}_L\ket{\downarrow}_R$ [see Fig.~\ref{fig:real_space}(c)], in accordance with the twofold degeneracy of the ES. Furthermore, as shown in Fig.~\ref{fig:real_space}(d), these two ground states also have gapless edge spin excitations: The spin sector $S_z=\pm 1$ exhibits degenerate ground states of the form $\ket{\uparrow}_L\ket{\uparrow}_R$ or $\ket{\downarrow}_L\ket{\downarrow}_R$, respectively. Therefore, the edge state manifold for a finite-size system, including both the $S_z=0$ and $S_z=\pm 1$ sectors, is fourfold degenerate. 

Let us now compare this degeneracy with the well-known case of the SSH model including the Hubbard interaction $U$, i.e., the SSH-Hubbard model \cite{Gurarie2011,Manmana2012,Yoshida2014,Wang2015,Ye2016,Sbierski2018,Barbiero2018,Le2020}. In the limiting case $U=0$, the SSH topological ground state exhibits a sixfold degeneracy composed by the states $\ket{\uparrow}_L\ket{\downarrow}_R, \ket{\downarrow}_L\ket{\uparrow}_R,  \ket{\uparrow\downarrow}_L\ket{0}_R, \ket{0}_L\ket{\uparrow\downarrow}_R$ in the $S_z=0$ sector, and $\ket{\uparrow}_L\ket{\uparrow}_R, \ket{\downarrow}_L\ket{\downarrow}_R$ for $S_z=\pm 1$. However, a finite $U$ results in an energy penalty for the states $\ket{\uparrow\downarrow}_L\ket{0}_R, \ket{0}_L\ket{\uparrow\downarrow}_R$, which become gapped. Hence, the interaction-induced BOW phase exhibits the same edge state manifold as a static dimerized model with on-site interaction $U$.

{\textit{Solitonic bulk excitations}}.\textemdash Another interesting aspect of the BOW in a finite chain is related to the interplay between spontaneous symmetry breaking and topology when bulk excited states are considered. In noninteracting topological insulators with static dimerizations, such as the SSH model, the excited bulk states are described as gapped modes carrying charge or spin on top of a background with a fixed sector of the dimerization. In contrast, here the dimerization is spontaneously induced via interactions, allowing for solitonic excitations interpolating between the two possible dimerization patterns. We note that, although topological defects also appear in the presence of phononic degrees of freedom~\cite{SSH1979,gonzalez_20a,gonzalez_20b}, here we observe them for single species through the frustration-induced spontaneous symmetry-breaking mechanism. We focus on bulk spin excitations of the topological BOW at half filling, as the charge gap is significantly larger \cite{Ejima2007}. 

\begin{figure}[t]
\includegraphics[width=\columnwidth]{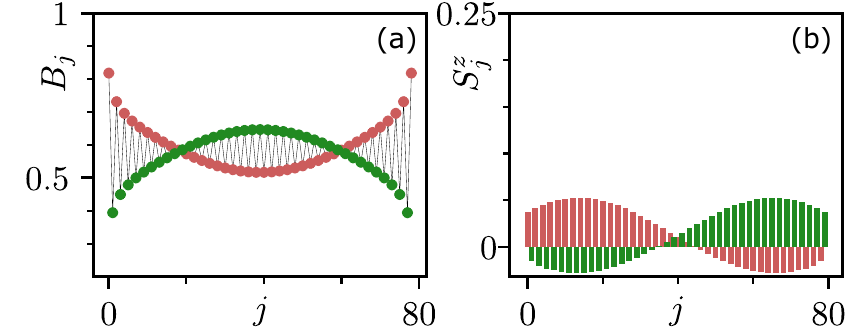} 
\caption{Finite DMRG results for $U=4t$, $V=2.14t$, $L=80$, and $\chi_\text{max}=1200$, with $S_z=+1$. (a) Solitonic profile in the bond order. (b) Local magnetization profile exhibiting a delocalized spin excitation.}
\label{fig:soliton}
\end{figure}

Importantly, such excitations also represent a route to obtain the topological sector of the BOW in a finite chain without relying on a pinning mechanism at the borders. This is what is shown in Fig.~\ref{fig:soliton}(a), where one can observe the first bulk spin excitation ($S_z=+1$) in the trivial BOW phase. We observe the solitonic domain walls interpolating between the trivial dimerization (left and right borders) to the topological one (central region). Notice that, as shown in Fig.~\ref{fig:soliton}(b), this corresponds to a delocalized soliton picture and thus this static solution is expected to be mobile; the DMRG solution corresponds to the minimum of a soliton band in the spin sector. The latter is reminiscent of a Peierl's mechanism with quantum phonons, but in the present case the solitons are generated by the same fermionic interactions. 

{\textit{Experimental proposal with ultracold dipolar gases}}.\textemdash
Hamiltonian~\eqref{eq:hamiltonianfree} can be simulated using a spin mixture in a dipolar Fermi gas of highly magnetic atoms. Since the emergence of the BOW phase is a rather general phenomenon, it can be experimentally investigated using various platforms and under realistic parameters. As an example, we consider lattice-confined fermionic erbium \cite{Patscheider2020} in a rectangular 3D lattice with spacings $(\Delta x,\Delta y,\Delta z)=(266, 266, 532)\ \text{nm}$ and lattice depths $(E_x, E_y, E_z)=(19,40,80)E_{\text{rec}}$. This results in tunneling rates $(t_x\equiv t,t_y,t_z)=(12.5,0.5, 0.001)\text{Hz}$, realizing the required effective 1D chains. Here, $E_{\text{rec}}$ is the photon recoil energy. The states $\ket{\uparrow}$ and $\ket{\downarrow}$ can be mapped into the two lowest Er Zeeman states. 
We find that the BOW phase, i.e., $U/t\sim 4$ and $V/t\gtrsim 2$, can be realized in the experiment with realistic parameters that allow us to match these conditions. For the above lattice parameter and a scattering length of $a_s=20a_0$ between $\ket{\uparrow}$ and $\ket{\downarrow}$, we indeed calculate $U=55\ \text{Hz}$ and $V=28\ \text{Hz}$. Notice that higher-order terms in the Hamiltonian, such as dipolar interactions beyond nearest neighbors~\cite{Barbiero2014}, density-induced tunneling~\cite{Jurguensen2014}, or spin-dependent dipolar terms, do not destabilize the BOW phases and its topological phases (see Supplemental Material \cite{SM} for an extended discussion).

Magnetic atoms lend themselves very well to all the Hamiltonian manipulation and engineering techniques developed with alkali atoms. This includes preparation of Mott states, spin manipulation, high-resolution imaging, and local control made accessible via microscopic techniques~\cite{Gross2021}. In addition, the rich atomic spectrum, distinctive of lanthanides~\cite{Norcia2021nof}, allows for new types of ultrafast control of the spin dynamics via optical manipulation based, e.g., on clock-type optical transitions~\cite{Petersen2020,Patscheider2021}.

\textit{Conclusions.}\textemdash {We showed that the BOW induced by frustration between competing couplings has a nontrivial topological sector in the presence of chiral symmetry. To this aim, we analyzed the BOW of the extended Fermi-Hubbard model. We revealed its topological nature by finding a nonzero string order correlator and a degenerate entanglement spectrum. We then discussed strategies to stabilize the topological sector in finite-size systems. The methods proposed in this Letter are general and can be used to analyze the topology of chiral-symmetric BOW phases induced by frustration, which are encountered in very diverse strongly-correlated quantum systems.
Finally, we also designed a realistic experimental scheme involving magnetic atoms trapped in an optical lattice where the topological BOW phase can be realized. The latter paves the way towards an efficient quantum simulation of topological phases in many-body quantum systems.} 

The ICFO group acknowledges support from ERC AdG NOQIA; Agencia Estatal de Investigación (R\&D Project No. CEX2019-000910-S, funded by MCIN/ AEI/10.13039/501100011033, Plan National FIDEUA PID2019-106901GB-I00, FPI, QUANTERA MAQS PCI2019-111828-2, QUANTERA DYNAMITE PCI2022-132919, Proyectos de I+D+I “Retos Colaboración” QUSPIN RTC2019-007196-7), MCIN via European Union NextGenerationEU (PRTR); Fundació Cellex; Fundació Mir-Puig; Generalitat de Catalunya through the European Social Fund FEDER and CERCA program (AGAUR Grant No. 2017 SGR 134, QuantumCAT \ U16-011424, co-funded by ERDF Operational Program of Catalonia 2014-2020); EU Horizon 2020 FET-OPEN OPTOlogic (Grant No. 899794); National Science Centre, Poland (Symfonia Grant No. 2016/20/W/ST4/00314); and European Union’s Horizon 2020 research and innovation programme under the Marie-Sk\l odowska-Curie Grant Agreements No. 101029393 (STREDCH) and No. 847648 (“La Caixa” Junior Leaders fellowships ID100010434: LCF/BQ/PI19/11690013, LCF/BQ/PI20/11760031, LCF/BQ/PR20/11770012, LCF/BQ/PR21/11840013). S.J.-F. acknowledges financial support from MCIN/AEI/10.13039/501100011033 and FSE “El FSE invierte en tu futuro” (reference code BES-2017-082118). A.D. further acknowledges the financial support from a fellowship granted by la Caixa Foundation (ID 100010434, fellowship code LCF/BQ/PR20/11770012). D.G.-C. is supported by the Simons Collaboration on Ultra-Quantum Matter, which is a grant from the Simons Foundation
(651440, P.Z.). The Innsbruck group acknowledges support from a QuantERA grant MAQS from the Austrian Science Fund (FWF No. I4391-N). DMRG calculations were performed using the TeNPy Library \cite{tenpy}. 

\bibliographystyle{apsrev4-2}

\end{document}


\title{Supplemental Materials: Revealing the topological nature of the bond order wave in a strongly correlated quantum system}

\author{Sergi Juli\`a-Farr\'e}
\email{sergi.julia@icfo.eu}
\affiliation{ICFO - Institut de Ciencies Fotoniques, The Barcelona Institute of Science and Technology, Av. Carl Friedrich Gauss 3, 08860 Castelldefels (Barcelona), Spain}
\author{Daniel González-Cuadra}
\affiliation{ICFO - Institut de Ciencies Fotoniques, The Barcelona Institute of Science and Technology, Av. Carl Friedrich Gauss 3, 08860 Castelldefels (Barcelona), Spain}
\affiliation{Center for Quantum Physics, University of Innsbruck, 6020 Innsbruck, Austria}
\affiliation{Institut f\"ur Quantenoptik und Quanteninformation, \"Osterreichische Akademie der Wissenschaften, Technikerstra{\ss}e 21a, 6020 Innsbruck, Austria}
\author{Alexander\,Patscheider}
\affiliation{Institut f\"ur Experimentalphysik, Universit\"at Innsbruck, Technikerstra{\ss}e 25, 6020 Innsbruck, Austria}
\author{Manfred\,J.\,Mark}
\affiliation{Institut f\"ur Quantenoptik und Quanteninformation, \"Osterreichische Akademie der Wissenschaften, Technikerstra{\ss}e 21a, 6020 Innsbruck, Austria}	
\affiliation{Institut f\"ur Experimentalphysik, Universit\"at Innsbruck, Technikerstra{\ss}e 25, 6020 Innsbruck, Austria}
\author{Francesca\,Ferlaino}
\affiliation{Institut f\"ur Quantenoptik und Quanteninformation, \"Osterreichische Akademie der Wissenschaften, Technikerstra{\ss}e 21a, 6020 Innsbruck, Austria}	
\affiliation{Institut f\"ur Experimentalphysik, Universit\"at Innsbruck, Technikerstra{\ss}e 25, 6020 Innsbruck, Austria}
\author{Maciej Lewenstein}
\affiliation{ICFO - Institut de Ciencies Fotoniques, The Barcelona Institute of Science and Technology, Av. Carl Friedrich Gauss 3, 08860 Castelldefels (Barcelona), Spain}
\affiliation{ICREA, Pg. Llu\'is Companys 23, 08010 Barcelona, Spain}
\author{Luca Barbiero}
\affiliation{Institute for Condensed Matter Physics and Complex Systems,
DISAT, Politecnico di Torino, I-10129 Torino, Italy}
\affiliation{ICFO - Institut de Ciencies Fotoniques, The Barcelona Institute of Science and Technology, Av. Carl Friedrich Gauss 3, 08860 Castelldefels (Barcelona), Spain}
\author{Alexandre Dauphin}
\email{alexandre.dauphin@icfo.eu}
\affiliation{ICFO - Institut de Ciencies Fotoniques, The Barcelona Institute of Science and Technology, Av. Carl Friedrich Gauss 3, 08860 Castelldefels (Barcelona), Spain}

\renewcommand{\theequation}{S\arabic{equation}}
\renewcommand{\thesection}{S\Roman{section}}
\renewcommand{\thefigure}{S\arabic{figure}}
\renewcommand{\bibnumfmt}[1]{[S#1]}
\renewcommand{\citenumfont}[1]{S#1}

\maketitle

\section{Pinning Hamiltonian for finite DMRG}
We use the pinning Hamiltonian 
\begin{equation}
H_{p}=2\delta\left(\sum_{j<j_0}(-1)^j B_j+ \sum_{ j\geq L-j_0}(-1)^j B_j\right)+V(\hat{n}_0+\hat{n}_{L-1}),
\end{equation}
 where $\delta$ is the strength of the pinning applied to the first and last $j_0$ bonds of the chain. For our calculations, we set $j_0=2$ ($H_p$ only acts on the two first and last bonds) and $|\delta|=0.5t$. For $\delta > 0$ ($\delta <0$) the topological (trivial) BOW is favored, as shown in Figs.~3(a)-(b) of the main text. The $V$ boundary term prevents the accumulation of unwanted charges at the edges caused by the sharp boundary conditions of the dipolar term in Hamiltonian (1) of the main text. In an experiment, a similar term would be present due to the fact that even though the edge sites would not be allowed to tunel outside of the chain they would still feel the dipolar interaction with surrounding atoms.
 
\section{Estimation of the MI-BOW BKT transition point}

Here we discuss the estimation of the MI-BOW and BOW-CDW transition points in the phase diagram of Fig.1(a) of the main text. The BOW-CDW transition can be accurately determined at $V_{\text{BOW-CDW}}\simeq 2.16t$ already from the behaviour of the BOW order parameter, as shown in the main text. However, the BOW order parameter remains finite in the MI due to Berezinskii-Kosterlitz-Thouless nature of the MI-BOW transition, with an exponentially slow gap opening, that leads to a divergence in the correlation length close to the transition point. This prevents a precise numerical treatment of this transition even with the iDMRG method. In short, this algorithm grows the size of the system at each iteration by assuming translational invariance of a given unit cell until reaching a fixed point at a large system size, which is assumed to be already in the thermodynamic limit. Close to the BKT transition the iDMRG does not find such a fixed point, and the state changes even for large system sizes due to the diverging correlation length.

\begin{figure}[h]
\includegraphics{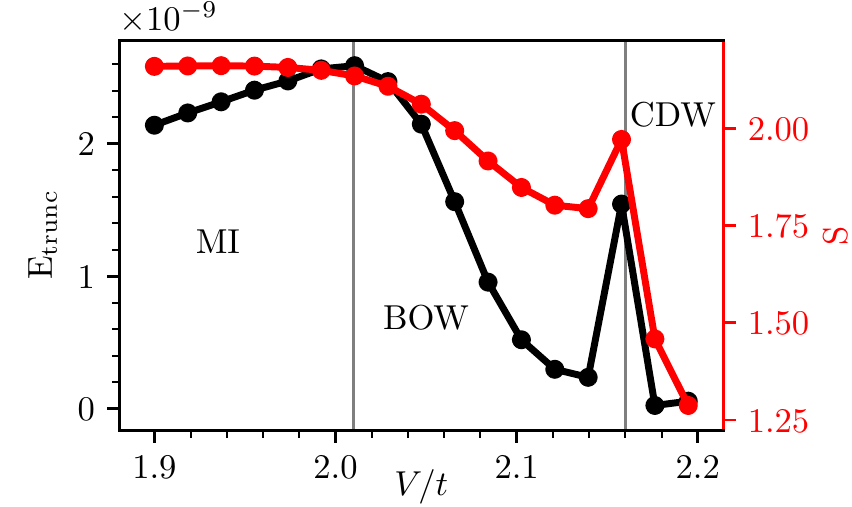}
\caption{Evolution of the truncation error $E_\textrm{trunc}$ and entanglement entropy $S$ of the iDMRG ground state for a fixed $U/t=4$ as a function of $V/t$. Here the iDMRG is iterated until a system size $L=20\ 000$ and considering a large bond dimension $\chi_{\textrm{max}}=3\ 000$.} 
\label{fig:line_trunc}
\end{figure}

However, the MI-BOW transition point can be estimated precisely by studying the lack of convergence of the iDMRG algorithm at a large bond dimension and system size. The latter can be analyzed with the behavior of both the entanglement entropy and the truncation error in the iDMRG ground state, as shown in Fig.~\ref{fig:line_trunc}. When entering the BOW phase, there is a change in the monotony of the truncation error of the iDMRG ground state. This is also accompanied by a reduction of the entanglement entropy, which indicates a reduction of the correlation length. From the results of this Fig.~\ref{fig:line_trunc}, we thus estimate  $V_\text{{MI-BOW}}\gtrsim 2.01t$, in agreement with the fact that in the thermodynamic limit there are logarithmic corrections to the finite-size estimates $V_\text{MI-BOW}\simeq 1.88t$~\cite{Ejima2007,Sengupta2002,Sandvik2004}, which shift the transition point to larger values of $V$. Such logarithmic corrections were modelled in the more recent finite-size extrapolations of Ref.~\cite{Dalmonte2015}, leading to $V_\text{{MI-BOW}}\sim 2.08t$.
\section{Entanglement entropy of the edge sites}
\begin{figure}[h]
\includegraphics{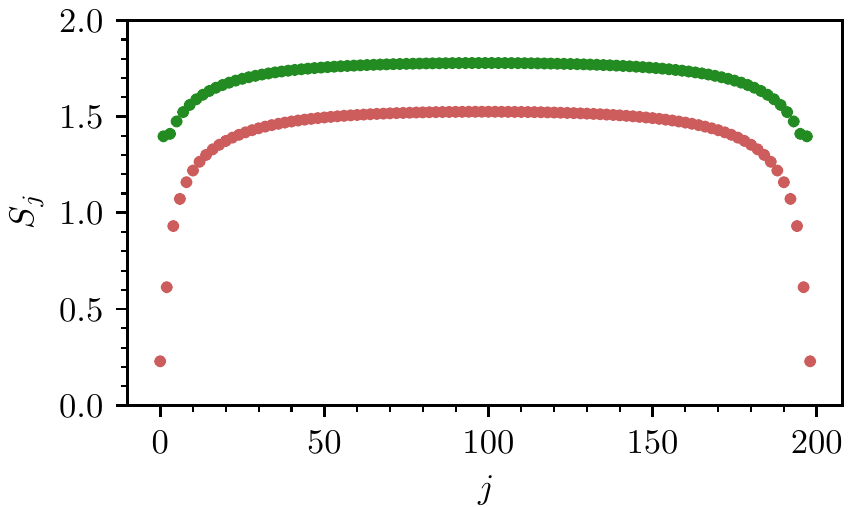} 
\caption{Entanglement entropy for a bipartite cut as a function of the $j$-th site in which the cut is performed. We use the same Hamiltonian parameters as in Fig.~3 of the main text, for which the system is in the topological sector of the BOW phase. Red (green) colors are used for even (odd) sites.}
\label{fig:entropy}
\end{figure}

Figure~\ref{fig:entropy} shows the entanglement entropy for different curs performed at each site of the chain for the case in which the system realizes a topological BOW phase. One observes a huge drop of the entropy for the cuts corresponding to the first/last sites being isolated from the rest of the chain. The low entanglement between these sites and the rest of the chain justifies the approximation of the state of the system as $\ket{\Psi}\simeq\ket{\Psi}_{\rm{bulk}}\ket{\Psi}_{\rm{edges}}$. Here $\ket{\Psi}_{\rm{edges}}=\ket{\cdot}_L\ket{\cdot}_R$, where $\ket{\cdot}_{L(R)}$ represents the quantum state of the first (last) site of the chain, and $\ket{\Psi}_{\rm{bulk}}$ the state of the bulk.

\section{Robustness of the BOW phase against extra Hamiltonian terms}

\subsection{Effect of density-induced tunneling}
In the experimental setup, one expects a small density-induced tunneling \cite{Jurguensen2014} $-\delta t\sum_{\langle ij\rangle,\sigma}(\cdop_{i,\sigma}\cop_{j,\sigma}(\hat{n}_{i,\bar{\sigma}}+\hat{n}_{j,,\bar{\sigma}})+\textrm{H.c.})$ of the order $\delta t/t\sim 10^{-2}$. Such a small term is expected to slightly renormalize the bare hopping strength.

\subsection{Effect of longer-range dipolar interactions}

In the Hamiltonian of Eq.(1) in the main text we have not considered dipolar interactions beyond nearest-neighbors as they decay as $1/r^3$. Nevertheless, the full dipolar repulsive potential has been shown to stabilize the BOW in a larger window of interactions \cite{Barbiero2014}.

\subsection{Effect of spin-dependent dipolar interactions}
Here we provide numerical evidence that the BOW of the Hamiltonian in Eq. (1) of the main text and its topological properties survive if the $V$ term is substituted by a spin-dependent dipolar interaction of the form $\sum_{\langle ij\rangle,\sigma \sigma '} V_{\sigma \sigma '}\hat{n}_{i,\sigma}\hat{n}_{j,\sigma '}$ with a moderate asymmetry between up and down species. To this aim we focus on the parameters $U/t=4$, $V_{\uparrow \downarrow}/t=2.14$, and consider $V_{\uparrow\uparrow}=0.9V_{\uparrow\downarrow}$, and $V_{\downarrow\downarrow}=1.1V_{\uparrow\downarrow}$. 

\begin{figure}[t]
\includegraphics{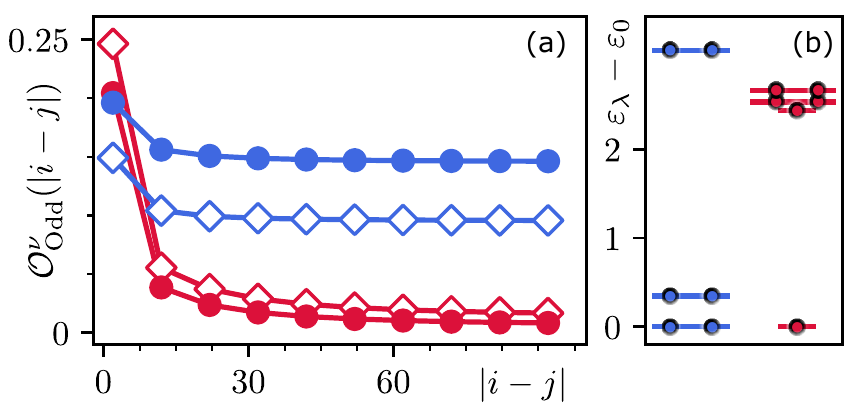}
\caption{Topological properties of the BOW at half filling for $U=4t$ and $V_{\uparrow \downarrow}=2.14t$, $V_{\uparrow\uparrow}=0.9V_{\uparrow\downarrow}$, and $V_{\downarrow\downarrow}=1.1V_{\uparrow\downarrow}$ obtained with iDMRG.  Red (blue) colors are used for the trivial (topological) sectors. (a) Long-range behavior of the odd string order parameters. Full circles (empty  squares) are used for the spin (charge) strings. (b) Entanglement spectrum.}
\label{fig:ent_so2}
\end{figure}

The iDMRG simulation of the resulting Hamiltonian at half filling leads to two degenerate ground states with $\Delta B=\pm 0.157$, in agreement with the spin-independent case (see Fig.~1(a) of the main text). The odd string order correlators for these two degenerate bulk ground states are shown in Fig.~\ref{fig:ent_so2}(a), and are in agreement with the spin-independent case (Fig.~2(a) of the main text). The entanglement spectrum are also shown in Fig.~\ref{fig:ent_so2}(b) and one can observe that the the one of the topological sector exhibits the same degeneracy as in the spin-independent case (Fig.~2(b) of the main text) and that the only appreciable difference is that the  breaking of the spin symmetry lifts the degeneracy of some excited entanglement energies in the trivial ground state.

\begin{figure}[t]
\includegraphics{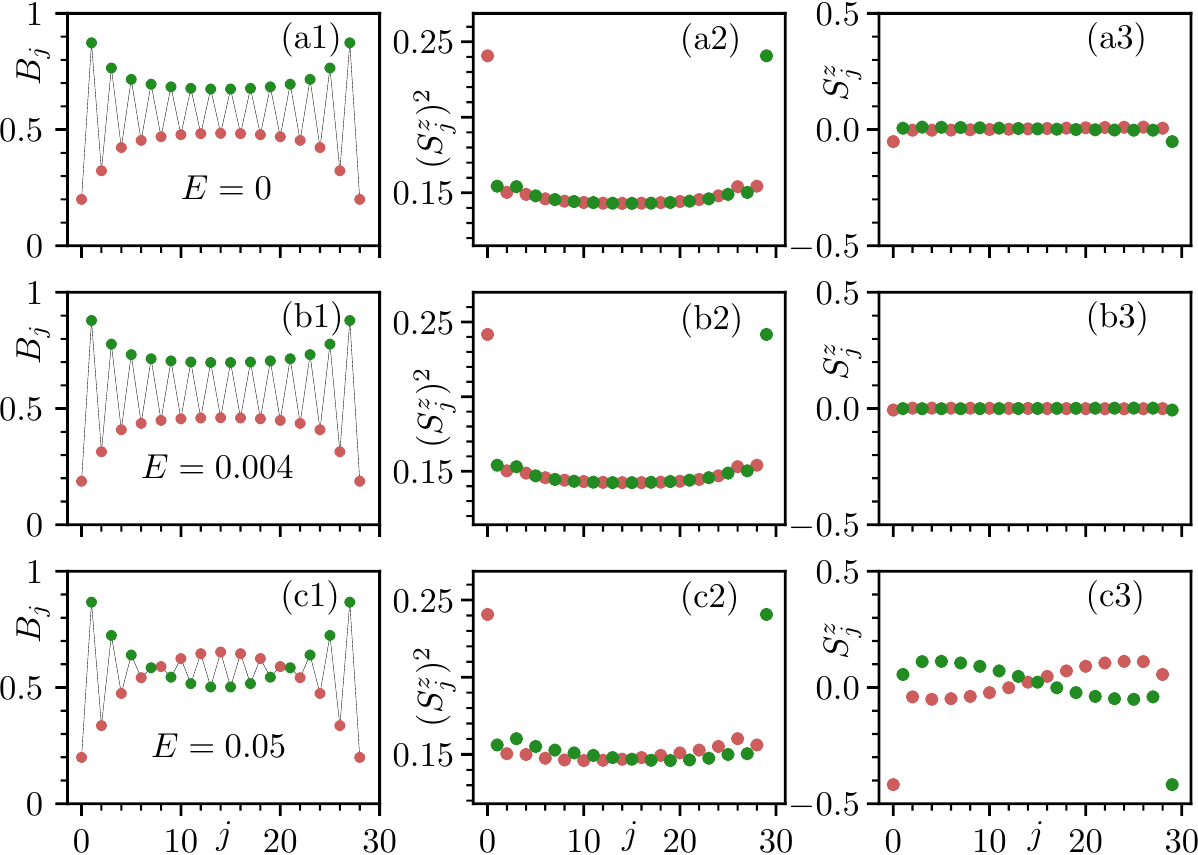}
\caption{Real-space properties of the BOW at half filling for $U=4t$ and $V_{\uparrow \downarrow}=2.14t$, $V_{\uparrow\uparrow}=0.9V_{\uparrow\downarrow}$, and $V_{\downarrow\downarrow}=1.1V_{\uparrow\downarrow}$ obtained with finite DMRG and $L=30$. The three rows (a),(b), and (c) correspond to the ground state and first two excited states, respectively. (a1)-(b1)-(c1) Bond order. (a2)-(b2)-(c2) Local spin-spin correlations.  (a3)-(b3)-(c3) Local spin polarization.}
\label{fig:SM_real}
\end{figure}

For studying the preservation of topological edge states we take a finite chain with $L=30$. A first observation is that the topological edge state manifold of the Hamiltonian in Eq.(1) of the main text presented a degeneracy between the $S_z=0,\pm 1$ sectors. With the spin-dependent repulsive potential that we are considering such a degeneracy is clearly broken, as the $S_z=+(-)1$ sector will pay more (less) energy penalty compared to the $S_z=0$ sector due to the fact that we are considering $V_{\downarrow\downarrow}<V_{\uparrow\downarrow}<V_{\uparrow\uparrow}$. Nevertheless, one still expects a twofold degeneracy within the $S_z=0$ sector because the spin-dependent potential does not break the inversion and chiral symmetries protecting the topological phase. This is indeed what is shown in Fig.\ref{fig:SM_real}, where we show the three lowest energy states of the topological sector of the BOW with $S_z=0$. One observes the presence of correlated edge states (signalled by a large value of $S_z^2$) in the two lowest states which are almost degenerate. The third state corresponds to a hybridization of the edge states of the $S_z=-1$ sector (which minimize the energy within the edge-state manifold) and a delocalized spin excitation in the bulk carrying $S_z=+1$ such that globally $S_z=0$. An important observation is that, for a given amount of asymmetry in $V_{\sigma \sigma '}$, this third state would become the ground state of the system below a critical value of the spin gap in the bulk. Such spin gap remains finite in the thermodynamic limit but diminishes as the system size is increased. The latter leads to a constrain to the maximum system size that one could used in an experiment given by the amount of spin asymmetry in the repulsive potential.
 
\section{Calculation of Hubbard parameters}
For the experimental values of $U$, $V$, $t$ and $\delta t$ we first calculate the 3D Wannier functions of the lowest band for the given cubic lattice $\phi(r)=\phi^x(x)\phi^y(y)\phi^z(z)$ and then evaluate the following terms numerically ($i$ and $j$ denote neighboring lattice sites along $x$):
\begin{equation}
\begin{split}
U = &\frac{4\pi\hbar a_s}{m}\int dr|\phi_i(r)|^4+ \\
&+\int dr dr'|\phi_i(r)|^2V_{\rm DDI}(r,r',\Theta)|\phi_i(r')|^2
\end{split}
\end{equation}
\begin{equation}
V_{\uparrow\downarrow} = \int dr dr'|\phi_i(r)|^2V_{\rm DDI}(r,r',\Theta)|\phi_j(r')|^2
\end{equation}
\begin{equation}
t = -\int dr \phi_i^*(r)(-\frac{\hbar^2\nabla^2}{2m}+V(r))\phi_j(r)
\end{equation}
\begin{equation}
\begin{split}
\delta t = &-\frac{4\pi\hbar a_s}{m}\int dr|\phi_i(r)|^2\phi_i^*(r)\phi_j(r)+ \\
&-\int dr dr'|\phi_i(r)|^2V_{\rm DDI}(r,r',\Theta)\phi^*_i(r')\phi_j(r')
\end{split}
\end{equation}
using
\begin{equation}
V_{\rm DDI}(r,r',\Theta) = \frac{\mu_0\mu_\uparrow\mu_\downarrow}{4\pi}\frac{1-3\cos^2(\Theta_{r-r'})}{|r-r'|^3}
\end{equation}
The numerical integration is done on a grid with 23x23x45 points on the full volume of a unit cell of the lattice centered on a lattice site for the DDI part of $U$, while for $V_{\uparrow\downarrow}$ and $\delta t$ a grid of 33x17x33 on the volume of two neighboring unit cells is used. The singularity is simply omitted, leading to a slight underestimation of the integral on the few percent level.